\documentclass{article}
\usepackage{epsfig}

\date{\today}

\begin{document}
\begin{center}

{\Large Rotating regular solutions in Einstein-Yang-Mills-Higgs theory }
\vspace{0.6cm}
\\
Vanush Paturyan$^{\ddagger}$,
 Eugen Radu$^{\dagger}$
and  D. H. Tchrakian$^{\dagger \star}$
\\
$^{\ddagger}${\small Department of
Computer Science, National University of Ireland Maynooth}
\\
$^{\dagger}${\small Department of
Mathematical Physics, National University of Ireland Maynooth,}
\\
$^{\star}${\small School of Theoretical Physics -- DIAS, 10 Burlington
Road, Dublin 4, Ireland }
\end{center}
\begin{abstract}
We construct new axially symmetric rotating solutions of
Einstein-Yang-Mills-Higgs theory. These globally regular
configurations possess a nonvanishing electric charge which equals
the total angular momentum, and zero topological charge,
representing a monopole-antimonopole  system rotating around the symmetry axis
through their common center of mass.
\end{abstract}

%%%%%%%%%%%%%%%%%%%%%%%%%%%%%%%%%%%%%%%%%%%%%%%%%%%%%%%%%%%%%%%%%%%%%%%%%%%%%%
%%%%%%                     Introduction
%%%%%%%%%%%%%%%%%%%%%%%%%%%%%%%%%%%%%%%%%%%%%%%%%%%%%%%%%%%%%%%%%%%%%%%%%%%%%%
\noindent{\textbf{Introduction.--~}} Rotation is an universal
phenomenon, which seems to be shared by all objects, at all possible
scales. For a gravitating Maxwell field, the Kerr-Newman black hole
solutions represent the only asymptotically flat configurations with
nonzero angular momentum. However, no regular rotating solution is
found in the limit of zero event horizon radius.

The inclusion of a larger (non Abelian) gauge group in the theory leads to
the possibility of regularising these configurations, as evidenced by the
Bartnick-McKinnon (BM) solution of the Einstein-Yang-Mills
(EYM) equations \cite{Bartnik:am}. However, to date no explicit example of
an asymptotically flat regular rotating solution with non Abelian matter
fields is known \cite{Volkov:2003ew}. Although
predicted perturbatively \cite{Brodbeck:1997ek}, no rotating
generalisations of the BM solution seem to exist
\cite{VanderBij:2001nm,Kleihaus:2002ee} \footnote{ Regular EYM
configurations with a nonzero angular momentum have been found only
in the presence of a negative cosmological constant \cite{Radu:2002rv}.
However, in this case the properties of the spherically symmetric
solutions differ substantially from those of the BM counterparts
\cite{Bjoraker:2000qd}.}.

The situation is more complicated in a spontaneously broken gauge
theory. As discussed in \cite{Heusler:1998ec,
VanderBij:2001nm,Volkov:2003ew} for
a Higgs field in the adjoint representation (the case considered in
this letter), the Julia-Zee dyons do no present generalisations with
a nonvanishing angular momentum. In fact the general result
presented in \cite{vanderBij:2002sq} proves that the angular
momentum of any regular solution with a nonvanishing magnetic charge
is zero~\footnote{Rotating black holes with a global magnetic charge
may exist and the first  set of such solutions have been recently
presented in \cite{Kleihaus:2004gm}.}. This however leaves open the
possibility of the existence of rotating Einstein-Yang-Mills-Higgs
(EYMH) solutions in the topologically trivial sector of the theory. We
have in mind  solutions described by an equal number of monopoles and
antimonopoles situated on the $z$-axis
with zero net magnetic charge like those in \cite{Kleihaus:2000hx} and
\cite{Kleihaus:1999sx}, gravitating and in flat space respectively; but
unlike the latter~\cite{Kleihaus:2000hx,Kleihaus:1999sx}, with nonzero
electric charge. Although the density of the
magnetic field is locally nonzero, the magnetic charge of these
configurations measured at infinity would vanish. This, in the
presence of an electric charge, results in nonzero angular momentum.

Despite the presence of some comments in the literature on the
possible existence of such solutions, no explicit construction has
been attempted.  Here we construct numerically the simplest example
of a regular rotating solution in a spontaneously broken gauge theory.
It represents an asymptotically flat, electrically charged
monopole-antimonopole (MA) system rotating around their common
center of mass. For a vanishing electric field, the solution reduces
to the static axially symmetric MA configurations discussed in
\cite{Kleihaus:2000hx}.

%%%%%%%%%%%%%%%%%%%%%%%%%%%%%%%%%%%%%%%%%%%%%%%%%%%%%%%%%%%%%%%%%%%%%%%%%%%%%%
%%%%%%                      The ansatz
%%%%%%%%%%%%%%%%%%%%%%%%%%%%%%%%%%%%%%%%%%%%%%%%%%%%%%%%%%%%%%%%%%%%%%%%%%%%%%
\noindent{\textbf{Axially symmetric ansatz and general
relations.--~}} Our study of the SU(2)-EYMH system is based upon the
action
\begin{equation}
S=\int \left ( \frac{R}{16\pi G} - \frac{1}{2} {\rm Tr} (F_{\mu\nu}
F^{\mu\nu}) -\frac{1}{4}{\rm Tr}(D_\mu \Phi D^\mu \Phi)
-\frac{1}{4}\lambda{\rm Tr}(\Phi^2 - \eta^2)^2
  \right ) \sqrt{-g} d^4x,
\end{equation}
with Newton's constant $G$, the Yang-Mills coupling constant $e$  
and Higgs self-coupling constant $\lambda$.

We consider  the usual Lewis-Papapetrou ansatz  \cite{Wald:rg} for a stationary,
axially symmetric spacetime with two Killing vector fields
$\partial/\partial  \varphi$ and $\partial/\partial t$.
In terms of the spherical coordinates $r,~\theta$ and $\varphi$, the
isotropic metric reads
\begin{equation}
ds^2=
  - f dt^2 +  \frac{m}{f} \left( d r^2+ r^2d\theta^2 \right)
           +  \frac{l}{f} r^2\sin^2\theta (d\varphi-\frac{\omega}{r} dt)^2
\ , \label{metric}
\end{equation}
where $f$, $m$, $l$ and $\omega$ are only functions of $r$ and $\theta$.

For the matter fields, we use a suitable parametrization of the
axially symmetric ansatz derived by Rebbi and Rossi
\cite{Rebbi:1980yi}, with a SU(2) gauge connection
\begin{equation}
\label{matter-ansatz}
 A_\mu dx^\mu =\vec A \cdot  d \vec r+A_t dt= \frac{1}{2er}
\left[ \tau _\phi
 \left( H_1 dr + \left(1-H_2\right) r d\theta \right)
 -\left( \tau_r H_3 + \tau_\theta \left(1-H_4\right) \right)
  r \sin \theta d\phi
+\left( \tau_r H_5 + \tau_\theta H_6 \right) dt \right] \ ,
\end{equation}
and a  Higgs field of the form
\begin{equation}
\Phi= \left(\Phi_1 \tau_r+\Phi_2 \tau_\theta\right)
\ . \end{equation}
The SU(2) matrices $(\tau_r,\tau_\theta,\tau_\phi)$
are defined in terms of the Pauli matrices
$\vec \tau = ( \tau_x, \tau_y, \tau_z) $ by
$\tau_r = \vec \tau \cdot
(\sin \theta \cos \phi, \sin \theta \sin  \phi, \cos \theta)$,
$\tau_\theta = \vec \tau \cdot
(\cos \theta \cos \phi, \cos \theta \sin  \phi, -\sin \theta)$,
$\tau_\phi = \vec \tau \cdot (-\sin  \phi, \cos  \phi,0)$.

The six gauge field functions $H_i$ and the two Higgs field function
$\Phi_i$ depend only on the coordinates $r$ and $\theta$ 
\footnote{The static MA solutions
discussed in \cite{Kleihaus:2000hx, Kleihaus:1999sx}
 have been obtained  for
a different parametrization of the same ansatz, imposed by a different choice of the
SU(2) matrices $(\tau_r,\tau_\theta,\tau_\phi)$.
Note that $H_5=H_6=0$ for static solutions.}.
We fix the
residual gauge degree of freedom by choosing the usual  gauge condition
$r\partial_rH_1-\partial_{\theta}H_2=0 $ 
\cite{Kleihaus:2002ee,Kleihaus:2004gm,Kleihaus:2000hx}.

Asymptotically flat, regular MA solutions are found by
imposing the boundary conditions
\begin{eqnarray}
\label{bc1a} f=m=l=1,~~\omega=0,~~H_1=H_3=0,~~H_2=H_4=-1,
\\
\nonumber
H_5=\gamma \cos \theta,~~H_6=\gamma \sin \theta,
~~\Phi_1=\eta \cos \theta,~~\Phi_2=\eta \sin \theta,
\end{eqnarray}
at infinity and
\begin{eqnarray}
\label{bc2a}
\nonumber
\partial_r f=\partial_r m=\partial_r l=\omega=0,~~H_1=H_3=0,~~H_2=H_4=1,
\\
\cos \theta\, \partial_r H_5-\sin \theta\,\partial_r H_6=0,~~ 
\sin \theta\, H_5+\cos \theta H_6=0,
\\
\nonumber
\cos \theta\, \partial_r\Phi_{1}-\sin \theta\,\partial_r
\Phi_{2}=0,~~ \sin \theta\, \Phi_{1}+\cos \theta\, \Phi_{2}=0,
\end{eqnarray}
at the origin. The functions $H_1,~H_3$ and the derivatives
$\partial_\theta f$, $\partial_\theta l$, $\partial_\theta m$,
$\partial_\theta \omega$, $\partial_\theta H_2$ and $\partial_\theta
H_4$ have to vanish for both $\theta=0$ and $\theta=\pi/2$.
 The other matter functions satisfy the boundary conditions
 $\partial_\theta H_5=H_6=\partial_\theta \Phi_1=\Phi_2=0$ on the
 $z-$axis ($\theta=0$) and  $H_5=\partial_\theta H_6=\Phi_1=\partial_\theta \Phi_2=0$
 on the $\rho-$axis ($\theta=\pi/2$).

The constants $\gamma,~\eta$ in (\ref{bc1a}) correspond to the
asymptotic magnitude of the electric potential and Higgs field,
respectively. The field equations imply the following expansion as
$r \to \infty$
\begin{eqnarray}
\label{exp1}
 f \sim 1-\frac{2M}{r}, ~~\omega \sim \frac{2J}{r^2},~~
H_5\sim \gamma \cos \theta (1-\frac{Q_e}{r}),~~H_6\sim \gamma \sin
\theta (1-\frac{Q_e}{r}).
\end{eqnarray}

The expression for the electric and magnetic charges derived by
using the 't Hooft field strength tensor  is
\begin{eqnarray}
\label{chargew} {\bf Q_e}=\frac{1}{4\pi}\oint_{\infty}dS_{\mu}Tr\{
\hat{\Phi}F_{\mu t} \}, ~~~ {\bf
Q_m}=\frac{1}{4\pi}\oint_{\infty}dS_{\mu}\frac{1}{2}\epsilon_{\mu
\nu \alpha} Tr\{\hat{\Phi}F_{\nu \alpha}\},
\end{eqnarray}
where $\hat{\Phi}=\Phi/|\Phi|$.   As implied by the asymptotic
behavior (\ref{bc1a}), (\ref{exp1}) these configurations carry zero
net magnetic charge, ${\bf Q_m}=0$ (although locally the magnetic
charge density is nonzero) and a nonvanishing electric charge ${\bf
Q_e}=\gamma Q_e$. Therefore they will present a magnetic dipole moment
$C_m$ which  can be obtained from the asymptotic form of the
non-abelian gauge field, after choosing a gauge where the Higgs
field is asymptotically constant $\Phi \to \tau_3$
\cite{Kleihaus:1999sx}
\begin{eqnarray}
\label{mon}
\vec A \cdot  d \vec r= C_m\frac{\sin^2 \theta}{r}\frac{\tau_3}{2} d\varphi.
\end{eqnarray}
The mass $M$ of these regular solutions is obtained form the
asymptotic expansion (\ref{exp1}) or equivalently from
$M=-\int (2 T_t^t-T_{\mu}^{\mu}) \sqrt{-g} dr
d\theta d\varphi$ \cite{Wald:rg}.
The constant $J$  appearing in (\ref{exp1})
corresponds to the total angular
%%%%%%%%%%%%% Figure 1: lambda dependence %%%%%%%%%%%%%%%%%%%%%%
\newpage
\setlength{\unitlength}{1cm}

\begin{picture}(18,7)
\centering
\put(2,0.0){\epsfig{file=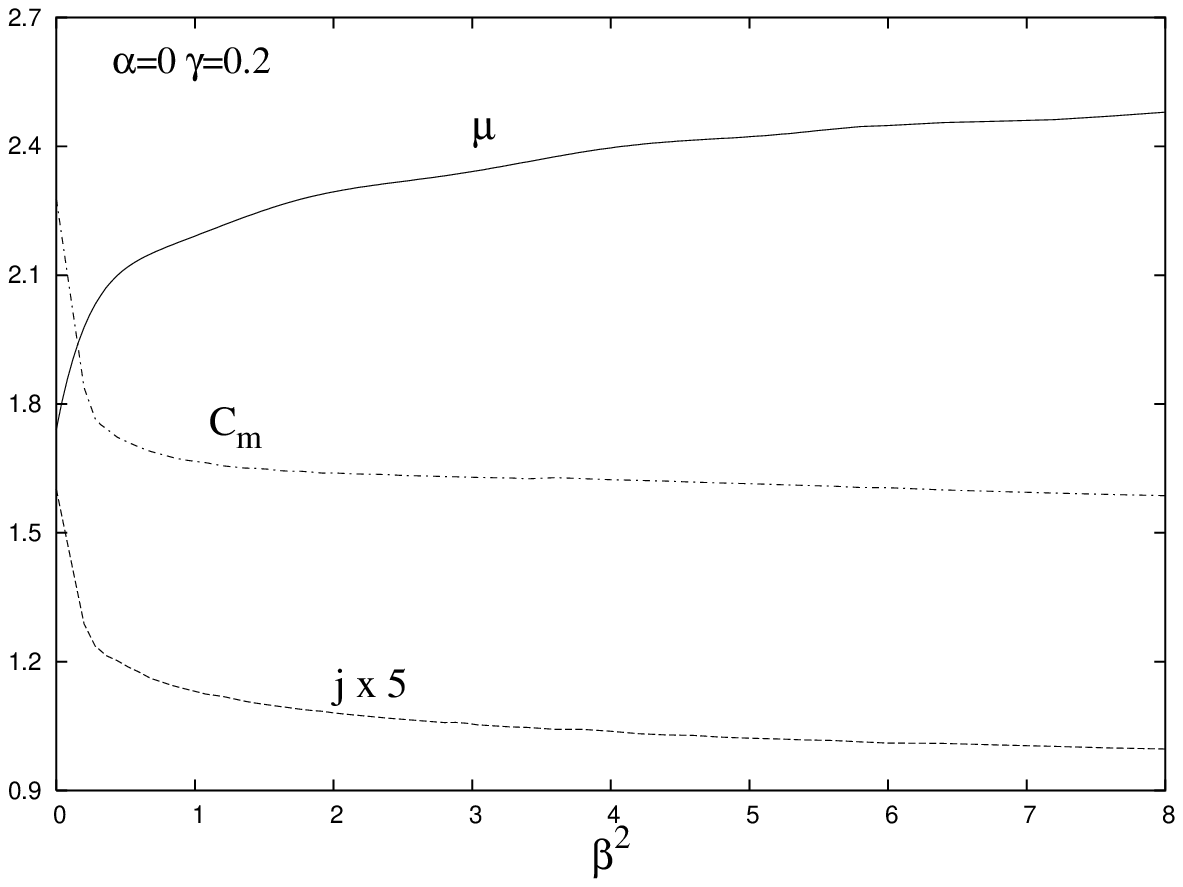,width=11cm}}
\end{picture}
\\
\\
{\small {\bf Figure 1.}
The mass, angular momentum and magnetic dipole moment are plotted
as a function of Higgs self-coupling constant $\beta^2$ for flat space
rotating MA solutions.}
\\
\\
 momentum of a solution which can
also be written as a volume integral
%\begin{eqnarray}
%\label{J}
$J= \int T_{\varphi}^{t}\sqrt{-g} dr d\theta
d\varphi$.
%\end{eqnarray}
As proven in \cite{VanderBij:2001nm}, another form of this expression,
in terms of asymptotics of the gauge potentials, is
\begin{eqnarray}
\label{totalJ} J &=&\oint_{\infty}dS_{\mu} 2Tr\{WF^{\mu t} \} ,
\end{eqnarray}
(with  $W=A_{\varphi}-\tau_z/2$), which, from the asymptotic expression
(\ref{exp1}) is just the electric charge, $J={\bf Q_e}$.
Introducing the dimensionless coordinate $x=r\eta e$ and the Higgs
field $\phi = \Phi/\eta$, the equations depend   on the coupling
constants $\alpha=\sqrt{4\pi G}\eta$   and
$\beta^2 =  \lambda/e^{2}$,
yielding the dimensionless mass and angular momentum $\mu=\frac{e}{4
\pi \eta}M$, $j=\frac{e  \eta^2}{4 \pi}J$.
%
%
%%%%%%%%%%%%%%%%%%%%%%%%%%%%%%%%%%%%%%%%%%%%%%%%%%%%%%%%%%%%%%%
%% NUMERICAL SOLUTIONS
%%%%%%%%%%%%%%%%%%%%%%%%%%%%%%%%%%%%%%%%%%%%%%%%%%%%%%%%%%%%%%%
%
\\
\noindent{\textbf{Numerical results.--}} We solve numerically the
set of twelve coupled non-linear elliptic partial differential
equations, subject to the above boundary conditions, employing a
compactified radial coordinate $\bar{x}=x/(1+x)$. As initial guess
we use the static MA regular solutions, corresponding to $\gamma=0$.
For any MA configuration, increasing $\gamma$ leads to rotating
regular solutions with nontrivial functions $H_5,~H_6$ and $\omega$.

For $\alpha=0$, we find rotating MA solutions in a flat spacetime
background. As remarked in \cite{Hartmann:2000ja}, for vanishing
 Higgs potential these solutions can be generated, from the
pure magnetic MA configuration ($\vec A,\Phi_0$) by using the
transformation  $\vec A \to \vec A$, $ \Phi \to \Phi_0 \cosh \xi$,
$A_t \to \Phi_0 \sinh \xi$ (no similar relation exists for
gravitating solutions although for small enough values of $\alpha$
the time component of the gauge field and the Higgs field are still
almost proportional). Their properties  can also be
deduced from the $\lambda=0$ MA configuration \cite{Hartmann:2000ja}.
To demonstrate the dependence  of the flat space MA rotating solutions
on the Higgs self-interaction,
we plot in Figure 1 the mass/energy, angular momentum and magnetic dipole
momentum as a function of $\beta$.
A similar qualitative picture is found for gravitating solutions.
However, all $\alpha \neq 0$ solutions presented here have no Higgs potential,
$\beta^2=0$.

When $\alpha$ is increased from zero, while keeping $\gamma$ fixed,
a branch of rotating solutions emerges from the flat spacetime
configurations. This branch ends at
a critical value $\alpha_{cr}$ which  depends on the
value of $\gamma$, the numerical errors increasing dramatically
for $\alpha>\alpha_{cr}$ for the solutions to be reliable. 
As $\alpha \to \alpha_{cr}$, the geometry remains
regular with no event horizon appearing, and, the mass and angular
momentum approach finite values.
Along this branch, the MA pair move closer to the origin and the mass,
angular momentum and magnetic dipole moment of the solutions decrease
to some limiting values
(see Figure 2).

As discused in \cite{Kleihaus:2000hx}, apart from the fundamental
branch, the static MA solutions admit also an infinite sequence of
excited configurations, emerging in the $\alpha \to 0$ limit (after
a rescaling)  from the spherically
%%%%%%%%%%%%%%%%%% Figure 2 alpha dependence %%%%%%%%%%%%%%%%%%%%%
\newpage
\setlength{\unitlength}{1cm}
\begin{picture}(18,7)
\centering
\put(2,0.0){\epsfig{file=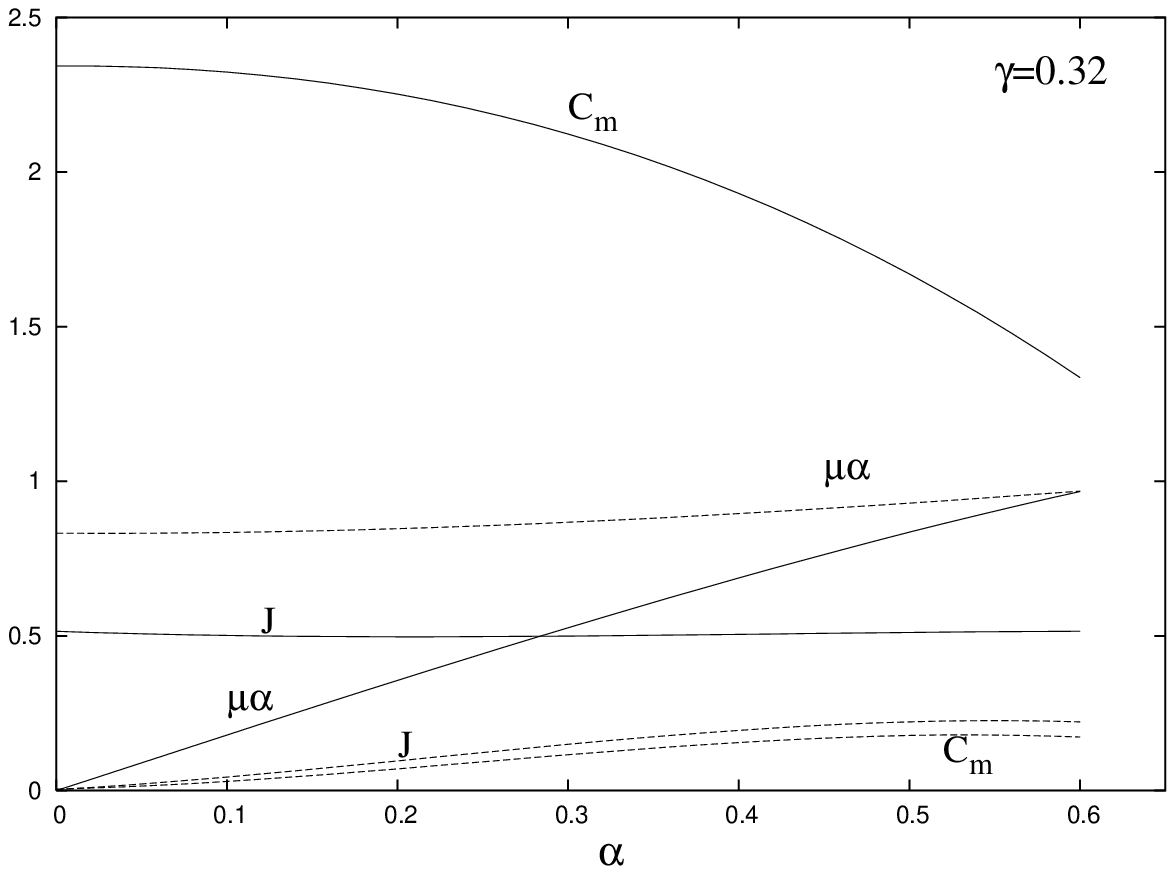,width=11cm}}
\end{picture}
\\
\\
{\small {\bf Figure 2.}  The scaled mass $\alpha \mu$,
the angular momentum $j$ and the magnetic dipole moment $C_m$
are shown as a function on $\alpha$
for a fixed value of the electric potential magnitude at infinity
$\gamma=0.32$. The solid and the doted lines
correspond to the fundamental and the second branch of solutions,
respectively.}
\\
\\
symmetric BM solutions
\footnote{For the fundamental branch solutions, $\alpha \to 0$
corresponds to $G \to 0$.}. The
lowest excited branch, originating  from the one-node BM solution,
evolves smoothly forward from $\alpha=0$ to $\alpha_{cr}$ where it
bifurcates with the fundamental branch.

 These static excited MA solutions present also rotating generalizations.
 For the
considered range of $\gamma$, we find that a second branch of
rotating MA solutions emerges at $\alpha_{cr}(\gamma)$, extending backwards
to $\alpha=0$.
As seen in Figure 2, although for a given $\gamma$ the mass of
the $\alpha_{cr}$ solutions is the same - within the limit of the
numerical accuracy,  other quantities such as angular momentum and
magnetic dipole moment present a discontinuity.
It is likely however that the complete picture is more complicated,
as in contrast to the static MA case, when $J\neq 0$ there is a black hole
solution with degenerate horizon for the regular solutions to merge into.
Thus, we expect more solutions to exist, representing a branch section
possibly
bending backwards in $\alpha$, and merging for some $\alpha_{m}$
into the extremal Kerr-Newman solution with $M^2=2 J^2$.
The numerical construction of such configurations presents
a considerable numerical challenge beyond the scope of the present work.

The excited solutions become infinitely heavy as $\alpha \to 0$
while the locations of the monopole and antimonopole
approach the
origin. The angular momentum/electric charge and the magnetic dipole
moment of the solutions vanish in the same limit.
The rescaling $x\to x \alpha$,
$\Phi \to \Phi/\alpha$ reveals that, similar to the static MA
solutions, the limiting solution on the upper branch is the first
spherically symmetric BM solution.
In this case, the limit $\alpha
\to 0$ corresponds to  $\eta \to 0$, for a nonzero value of $G$.
The
limiting value of the scaled mass $\hat{\mu}=\alpha \mu$ corresponds
also to the mass of the one-node BM solution, with $H_1=H_3=0$,
$H_2=H_4=w(r)$, $H_5=H_6=0$. Thus, no rotating limiting EYM solution
is found in this way. Without a Higgs field, the regularity
conditions force the electric potentials to vanish
identically~\footnote{Spherically symmetric EYM solutions with
$A_t \neq 0$ cannot exist \cite{bizon},
while the rotating black hole solutions have $\gamma=0$
\cite{Kleihaus:2002ee}, the electric field being supported by the
rotating event horizon contribution.}.

Indeed, for any value of $\alpha$, we could not find solutions with
an asymptotic magnitude of electric potential greater than that of
the Higgs field (i.e. $\gamma > 1$ after rescaling).
In this situation, similar to the case of dyon configurations
\cite{Hartmann:2000ja}, the asymptotic behavior of some gauge field
functions became oscillatory, failing to satisfy the required
boundary conditions.

In Figure 3 we show the mass, angular momentum and magnetic dipole moment
as a function of $\gamma$ for a fixed value of $\alpha$. Both fundamental
and second branch solutions are displayed.
These quantities increase always with $\gamma$ and stay finite
as $\gamma \to 1$.
%
%%%%%%%%%%%%% Figure 3: eta dependence %%%%%%%%%%%%%%%%%%%%%%
\newpage
\setlength{\unitlength}{1cm}

\begin{picture}(18,7)
\centering
\put(2,0.0){\epsfig{file=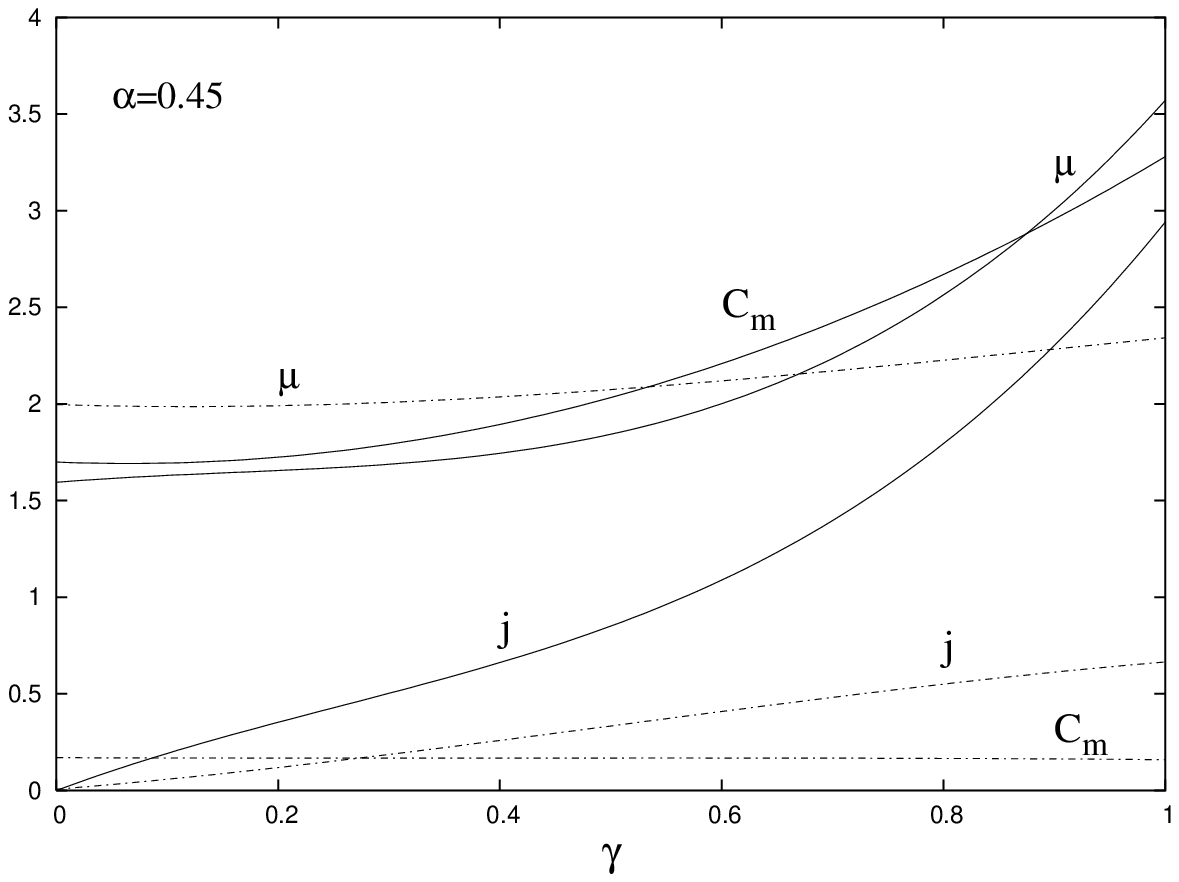,width=11cm}}
\end{picture}
\\
\\
{\small {\bf Figure 3.} The mass $\mu$, the angular momentum $j$
and the magnetic dipole moment $C_m$
are shown as a function on the magnitude of the electric potential at
infinity $\eta$ for a fixed value of $\alpha=0.45$. The solid and the
doted lines correspond to the fundamental and the second branch of
solutions, respectively.}
\\
\\
The rotating solutions share a number of
common properties with the purely magnetic MA configurations. The
modulus of the Higgs field possesses always two zeros at $\pm d/2$
on the $z-$symmetry axis, corresponding to the location of the
monopole and antimonopole, respectively.  Both $\vec A$ and $\Phi_a$
present a  shape similar to the static case. The energy density
$\epsilon = -T_{t}^t$ possesses maxima at $z=\pm d/2$ and a saddle
point at the origin, and presents the typical form exhibited in the
literature on MA solutions. A different picture is found for the
angular momentum density. As seen in Figure 4, the MA system rotates as a
single object and the $T_\varphi^t$-component of the energy momentum
tensor associated with rotation presents a maximum in the $z=0$ plane and
no local extrema at the locations of the monopole and the antimonopole.

Although we have restricted  the analysis  here to the simplest sets
of solutions, rotating MA configurations have  been found also
starting with excited MA branches with $A_t=0$. These solutions do
not possess counterparts in flat spacetime and their  $\alpha
\to 0$ limit corresponds always to (higher node-) BM solutions.
%
%%%%%%%%%%%%%%%%%%%%%%%%%%%%%%%%%%%%%%%%%%%%%%%%%%%%%%%%%%%%%%%%%%%%%%%%%%%%%%
%                         FURTHER DISCUSSIONS
%%%%%%%%%%%%%%%%%%%%%%%%%%%%%%%%%%%%%%%%%%%%%%%%%%%%%%%%%%%%%%%%%%%%%%%%%%%%%%
%
\noindent{\textbf{Further remarks.--~}} We have presented here the
first set of globally regular solutions of EYMH theory possessing a
nonvanishing angular momentum.  These asymptotically flat
configurations carry mass, angular
 momentum, electric charge and no
net magnetic charge. The electric charge is induced by rotation and
equals the total angular momentum.

The excited rotating solutions do not possess a counterpart in flat
space, and their angular momentum vanishes in the no-Higgs field
limit, which corresponds to the BM configurations. The nonexistence
of a rotating generalisation of the BM solution can be viewed as a
consequence of the impossibility to obtain regular, electrically
charged nonabelian solutions without a Higgs field.

The situation here differs from the electrically neutral gravitating
case~\cite{Kleihaus:2000hx}, where there is no available black hole
solution (e.g. Reissner-Nordstr\"om) for the fundamental branch to end in,
due to the absence of a global (magnetic) charge. Here by contrast we have
an electric charge, so one might expect the fundamental branch to end in
the corresponding rotating black hole, namely the Kerr-Newman solution.
Our numerical results indicate tentatively that this might well be the
case, as the metric functions seem to be decreasing towards zero with
$\alpha$ increasing beyond the bifurcation point $\alpha_{cr}$.
Unfortunately we cannot make this claim reliably here, because the
numerical accuracy of these results is not sufficiently good. The
complexity of this numerical task is beyond the scope of the present work,
and remains an outstanding matter to be disposed of in future.

Concerning the stability of these solutions, in the absence of a
topological charge, we expect them to be
unstable, similar to the electrically uncharged MA configurations.

%%%%%%%%%%%%% Figure 4: T34 %%%%%%%%%%%%%%%%%%%%%%
\newpage
\setlength{\unitlength}{1cm}

\begin{picture}(11,5.5)
\centering
\put(2,0.0){\epsfig{file=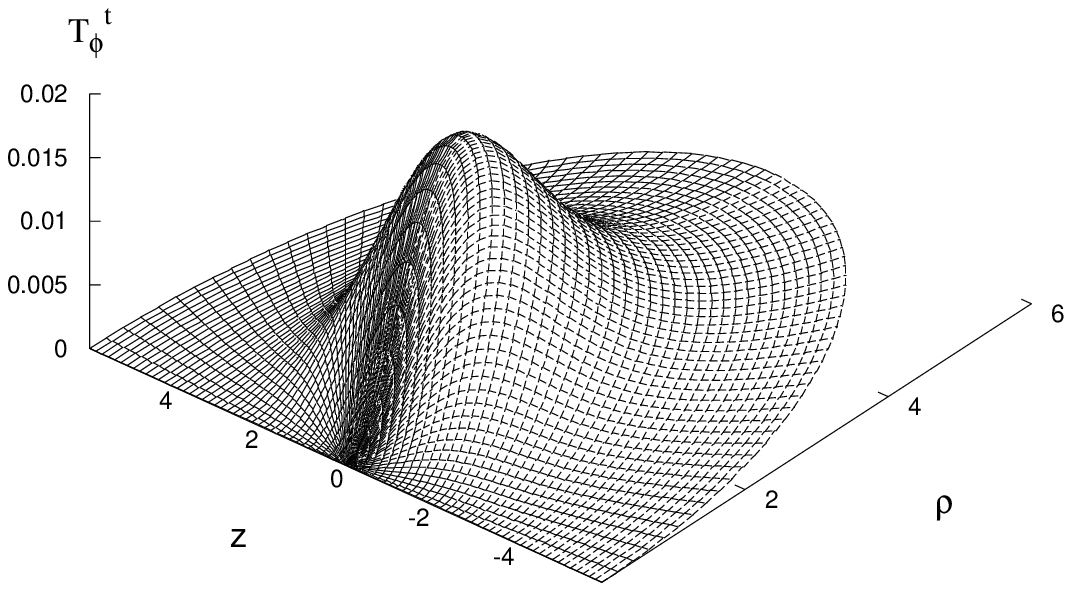,width=11cm}}
\end{picture}
\\
\\
{\small {\bf Figure 4.} The angular momentum density $T_{\varphi}^t$ is shown for
 a typical fundamental branch MA rotating solution, with $\alpha=0.42$,~$\gamma=0.26$.}
\\
\\
By including an integer $n$ (the winding number) in the ansatz
(\ref{matter-ansatz}),  rotating MA chains and rotating
vortex rings
can be found for $n>1$ . We expect also that EYMH theory possesses a
whole sequence of rotating solutions, obtained within the ansatz
(\ref{matter-ansatz}), for an asymptotic behaviour of the Higgs field
with $\Phi_1=\eta \cos (2k+1) \theta,~~\Phi_2=\eta (2k+1) \sin
\theta$, where $k$ is a positive integer (the static limit of these
solutions is discussed recently in \cite{Kleihaus:2004fh}). 
These solutions would possess again an angular momentum equal to the
electric charge but no net magnetic charge. The $\alpha \to 0$ limit of
the excited solutions will correspond to the recently discovered sequence
of EYM static axially symmetric configurations \cite{Ibadov:2004rt}.

Rotating MA black hole solutions, generalizing the static, axially
symmetric black holes with magnetic dipole hair \cite{Kleihaus:2000kv},
should exist as well. However, these solutions will not satisfy the
intriguing relation ${\bf Q_e}/J=1$ which is a unique property of the
regular configurations with zero topological charge.
\\
\\
{\bf Acknowledgement}
\\
This work was carried out in the framework of Enterprise--Ireland
Basic Science Research Project SC/2003/390.
%%%%%%%%%%%%%%%%%%%%%%%%%%%%%%%%%%%%%%%%%%%%%%%%%%%%%%%%%%%%%%%%%%%%%%%%%%%%%%

\end{document}